\documentclass[fleqn,10pt]{wlscirep}
\title{Programmable control of spin-wave transmission in a domain-wall spin valve}

\author[1]{Sampo J. Hämäläinen}
\author[2]{Marco Madami}
\author[1]{Huajun Qin}
\author[3]{Gianluca Gubbiotti}
\author[1,*]{Sebastiaan van Dijken}
\affil[1]{NanoSpin, Department of Applied Physics, Aalto University School of Science, P.O. Box 15100, FI-00076 Aalto, Finland}
\affil[2]{Dipartimento di Fisica e Geologia, Università di Perugia, I-06123 Perugia, Italy}
\affil[3]{Istituto Officina dei Materiali del CNR (CNR-IOM), Sede Secondaria di Perugia, c/o Dipartimento di Fisica e Geologia,
	Università di Perugia, I-06123 Perugia, Italy}

\affil[*]{sebastiaan.van.dijken@aalto.fi}

\begin{abstract}
Active manipulation of spin waves is essential for the development of magnon-based technologies. Here, we demonstrate programmable spin-wave filtering by resetting the spin structure of a pinned 90$^\circ$ N\'{e}el domain wall in a continuous CoFeB film with abrupt rotations of uniaxial magnetic anisotropy. Using phase-resolved micro-focused Brillouin light scattering and micromagnetic simulations, we show that broad 90$^\circ$ head-to-head or tail-to-tail magnetic domain walls are transparent to spin waves over a broad frequency range. In contrast, magnetic switching to a 90$^\circ$ head-to-tail configuration produces much narrower domain walls at the same pinning locations. Spin waves are strongly reflected by a resonance mode in these magnetic domain walls. Based on these results, we propose a magnetic spin-wave valve with two parallel domain walls. Switching the spin-wave valve from an open to a close state changes the transmission of spin waves from nearly 100\% to 0\% at the resonance frequency. This active control over spin-wave transport could be utilized in magnonic logic devices or non-volatile memory elements.
\end{abstract}
\begin{document}

\flushbottom
\maketitle
\thispagestyle{empty}

\section*{Introduction}
Wave-like computing based on spin waves has generated interest as a potential low-power and parallel computing alternative for conventional CMOS technologies\cite{CHU-15}. High group velocities\cite{YU-12}, long decay lengths\cite{LIU-18}, and the ability to tailor the wavelength of spin waves down to the nanoscale\cite{YU-16,VAN-16,STI-18} offer fascinating prospects for magnonics. Logic devices based on a Mach-Zehnder spin-wave interferometer have been proposed and realized\cite{SCH-08,LEE-08,KHI-10,CHU-14}. In this geometry, spin waves are launched into the interferometer branches and their phase or amplitude is controlled by an Oersted field\cite{SCH-08,LEE-08,KHI-10} or spin-wave current from a third magnetic terminal\cite{CHU-14}. Interference of spin-wave signals after manipulation determines the logic output. Other logic concepts exploit reprogrammable magnonic crystals\cite{KRA-14,CHU-17}. Here, spin-wave transmission is controlled actively by a spatial modulation of magnetic properties using, for instance, electric currents\cite{CHU-09} or optical pulses\cite{VOG-15}. Besides these dynamic approaches, field-induced toggling between ferromagnetic and antiferromagnetic states in magnetic stripe arrays has been shown to modify spin-wave transport in a non-volatile way\cite{TOP-10,GUB-18}. 

Non-collinear spin structures, such as magnetic domain walls, can also be exploited to control the amplitude or phase of spin waves\cite{HER-04}. It has been analytically calculated that spin-wave transport through an infinitely extended one-dimensional Bloch wall induces a finite phase shift without reflection\cite{BAY-05}. In contrast, interactions between spin waves and 180$^\circ$ N\'{e}el walls or magnetic domain walls in confined geometries are more complex. Dynamic stray fields in such domain walls reduce the transmission of spin waves if their wavelength exceeds the wall width\cite{MAC-10,WAN-APL-13}. Additionally, domain-wall resonances limit the transmission of spin waves at specific frequencies\cite{HAN-09,KIM-12,WAN-12,WAN-JAP-13}. This effect, known as resonant reflection, relates strongly to the spatial inhomogeneity of the effective magnetic field inside the wall. Thus, because of dynamic stray fields and resonance modes, narrow domain walls reflect spin waves more than broad walls. In many studies, the interaction between spin waves and magnetic domain walls is investigated as a new method to drive walls into motion\cite{HIN-11,YAN-11,WAN-APL-13,HAN-09,KIM-12,WAN-12,WAN-JAP-13}. It has been shown that the direction and velocity of wall motion depend strongly on the spin-wave transmission coefficient. For coefficients close to unity, transfer of angular momentum causes domain walls to move against the spin waves\cite{HIN-11,YAN-11}. In contrast, strong spin-wave reflection at domain walls induces forward domain-wall motion\cite{WAN-APL-13,HAN-09,KIM-12,WAN-12,WAN-JAP-13}. 

\begin{figure}[ht]
	\centering
	\includegraphics[width=\linewidth]{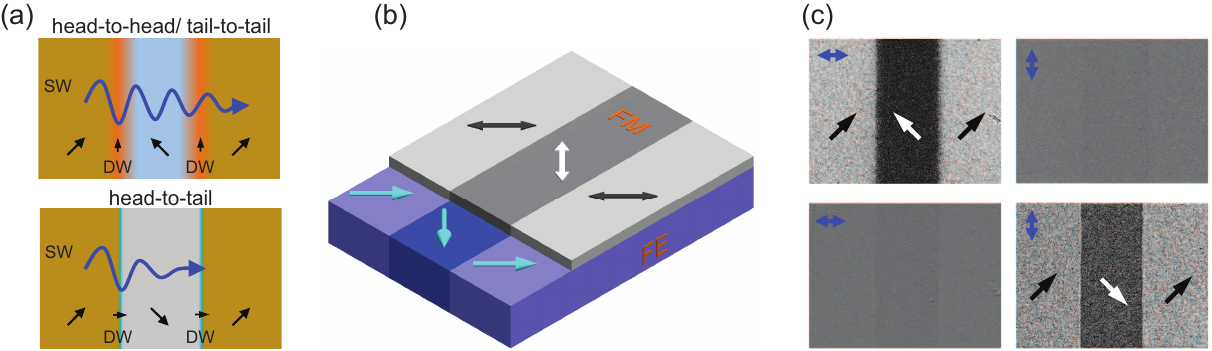}
	\caption{(a) Schematic of the magnetic spin-wave valve. The structure consists of two pinned magnetic domain walls in a ferromagnetic film separating stripe domains with uniaxial magnetic anisotropy. Because the anisotropy axis rotates by 90$^\circ$ at the domain boundaries, alternating head-to-head and tail-to-tail domain walls are initialized after the application of a magnetic field along the walls (top). These magnetic domain walls are broad and transparent to spin waves. The application of a perpendicular field results in the formation of much narrower head-to-tail domain walls (bottom). Spin waves are mostly reflected by these walls. Reversible magnetic switching in the central domain thus toggles the efficiency of spin-wave transmission. (b) Illustration of the experimental CoFeB/BaTiO$_3$ bilayer system. The domain pattern of the ferroelectric (FE) substrate is imprinted into the ferromagnetic (FM) film via strain transfer and inverse magnetostriction. Blue arrows and double-headed black and white arrows indicate the direction of ferroelectric polarization and the orientation of uniaxial magnetic anisotropy. (c) Magneto-optical Kerr microscopy images of the magnetization configuration in a CoFeB/BaTiO$_3$ bilayer after the application of a magnetic field along the domains walls (top) and perpendicular to the domain walls (bottom). The left and right images are taken with the magneto-optical contrast axis along the horizontal and vertical direction, respectively (blue arrows).}
	\label{fig:1}
\end{figure}

In this paper, we demonstrate the use of two parallel magnetic domain walls as magnetic spin-wave valve (Fig. \ref{fig:1}(a)). Periodic 90$^\circ$ rotations of uniaxial magnetic anisotropy firmly pin N\'{e}el-type walls in our system, preventing their motion under the action of propagating spin waves. Because of pinning, a magnetic field can reversibly transform a narrow 90$^\circ$ head-to-tail domain wall into a broad 90$^\circ$ head-to-head or tail-to-tail wall. Using phase-resolved micro-focused Brillouin light scattering ($\mu$-BLS) and micromagnetic simulations, we show that broad domain walls are transparent for spin waves over a wide frequency range. In contrast, spin waves are resonantly reflected by narrow domain walls. Reprogramming of the non-volatile domain-wall structure in a spin-wave valve with two parallel walls toggles the transport characteristics from fully transparent to a transmission coefficient of nearly 0\% at the resonance frequency. 

\section*{Experimental realization}

Interactions of spin-waves with 90$^\circ$ N\'{e}el walls have been imaged in square-shaped ferromagnetic microstructures with a Landau domain state\cite{BUE-06,PIR-15}. The head-to-tail domain walls in these experiments act as barriers for spin-wave transport. To realize active switching between such narrow domain walls and broader 90$^\circ$ walls with a head-to-head or tail-to-tail structure, we couple a 50 nm thick ferromagnetic CoFeB film to a ferroelectric BaTiO$_3$ substrate with regular ferroelastic stripe domains (Fig. \ref{fig:1}(b)). The combination of strain transfer and inverse magnetostriction in this bilayer system causes imprinting of magnetic stripe domains in the ferromagnetic layer\cite{LAH-11,LAH-12}. The magnetic anisotropy of the CoFeB film is uniaxial, reflecting the tetragonal symmetry of the underlying ferroelectric crystal, and the anisotropy axis rotates abruptly by 90$^\circ$ between domains. In a previous study, we showed that excitation of this system by an uniform microwave magnetic field results in the formation of standing spin waves within the domains and spin-wave emission from the anisotropy boundaries\cite{HAM-17}. Here, we focus on spin-wave transport through 90$^\circ$ N\'{e}el walls. Importantly, strong pinning of straight magnetic domain walls at the anisotropy boundaries enables us to switch between head-to-tail and head-to-head/tail-to-tail walls in the CoFeB film. We use an external in-plane magnetic field to switch between these non-volatile magnetization states. If the field is applied along the domain walls and turned off, the magnetization aligns in an alternating head-to-head and tail-to-tail configuration (Fig. \ref{fig:1}(c), top). The same process after rotating the in-plane field by 90$^\circ$ initializes straight head-to-tail domain walls instead (Fig. \ref{fig:1}(c), bottom). Both walls are of the 90$^\circ$ N\'{e}el type in zero magnetic field, but their width differs significantly\cite{FRA-12}. Scanning electron microscopy with polarization analysis (SEMPA), X-ray photoemission electron microscopy (XPEEM), and micromagnetic simulations on CoFeB/BaTiO$_3$ and related systems have previously shown that the width ($\delta$) of 90$^\circ$ head-to-head/tail-to-tail domain walls increases sharply with CoFeB film thickness ($t$) because of their large magnetostatic energy ($\delta\propto{t}$)\cite{FRA-14,CAS-15,LOP-17}. In contrast, the width of 90$^\circ$ head-to-tail domain walls is almost entirely determined by a competition between ferromagnetic exchange and the strength of magnetic anisotropy and, thus, it varies much less with $t$\cite{FRA-14}. Consequently, the difference in domain wall width is more pronounced in thick ferromagnetic films. In our experiments, we focus on 50-nm-thick CoFeB because it combines full imprinting of magnetic stripe domains (Figs. \ref{fig:1}(b) and 1(c)) with giant reprogramming of the domain wall width from 50 nm (head-to-tail wall) to 1.6 $\mu$m (head-to-head and tail-to-tail wall). Details on sample preparation are given in the Methods section.  

\section*{Programmable spin-wave transmission through a 90$^\circ$ magnetic domain wall}

\subsection*{Brillouin light scattering}

\begin{figure}[ht]
	\centering
	\includegraphics[width=\linewidth]{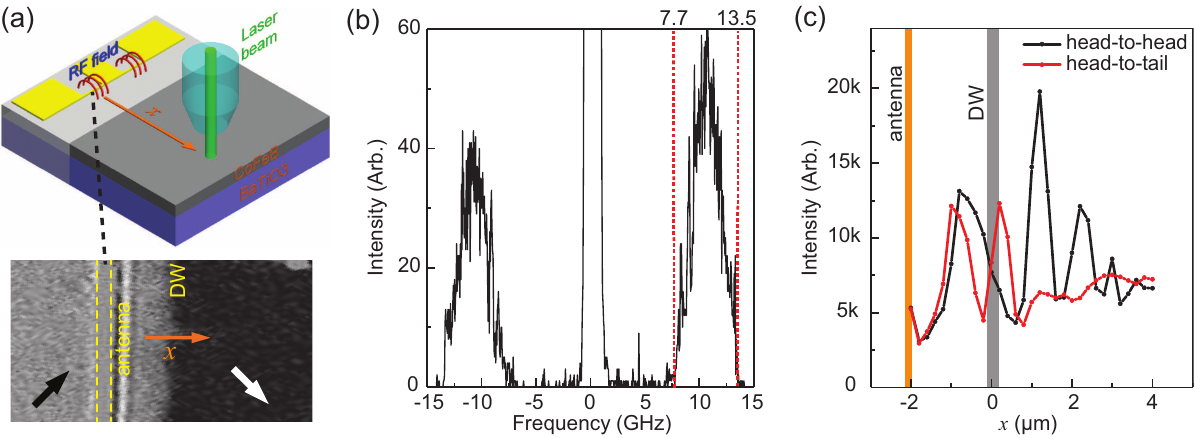}
	\caption{(a) Schematic and magneto-optical Kerr microscopy image of the $\mu$-BLS measurement geometry. A microwave antenna on top of the CoFeB film excites propagating spin waves and the transmission of these spin waves through a nearby head-to-head or head-to-tail domain wall is recorded by scanning the $\mu$-BLS laser beam across the wall (orange arrow). (b) Local BLS measurement as a function of frequency at a fixed location 500 nm from the antenna. (c) Phase-resolved $\mu$-BLS scans across the domain wall in the CoFeB film for a 90$^\circ$ head-to-head (black curve) and 90$^\circ$ head-to-tail (red curve) magnetization configuration. The excitation frequency is 10.85 GHz.}
	\label{fig:2}
\end{figure}

Dispersion relations of spin waves depend on the angle between the wave vector and direction of magnetization. In most spin-wave experiments, a magnetic bias field is used to saturate the magnetization of a sample either parallel or perpendicular to the wave vector. Because of strain-induced uniaxial magnetic anisotropy in the CoFeB film of our bilayer, the magnetization is uniform within the stripe domains in zero magnetic field. This enabled us to perform measurements without bias field and to study spin-wave transport through 90$^\circ$ N\'{e}el walls. In the remanent state, the CoFeB magnetization aligns along the uniaxial magnetic anisotropy axes of the stripe domains, i.e., at an angle of 45$^\circ$ with respect to the domain walls (Fig. \ref{fig:1}(c)). The type of magnetic domain wall was set before spin-wave characterization by either applying a parallel or perpendicular magnetic field. 

We employed $\mu$-BLS to measure spin-wave transmission through pinned magnetic domain walls. For the excitation of spin waves, we patterned 500-nm-wide microwave antennas on top of our sample using electron-beam lithography. The antennas are separated from the CoFeB film by an insulating TaO$_x$ layer and they are aligned parallel to a nearby domain wall (Fig. \ref{fig:2}(a), top). The microwave antennas excite spin waves over a broad range of wave vectors (the shortest wavelength is about 1 $\mu$m). In the experiments described below, the antenna edge and domain-wall center are separated by approximately 2 $\mu$m (Fig. \ref{fig:2}(a), bottom). We first optimized the $\mu$-BLS measurement sensitivity by performing a frequency scan in the range from 7.7 to 13.5 GHz at a fixed location. The data in Fig. \ref{fig:2}(b), recorded 500 nm from the antenna, show a maximum signal intensity at 10.85 GHz. We used this frequency to spatially resolve the profile of propagating spin waves before and after transmission through the 90$^\circ$ magnetic domain wall.

Figure \ref{fig:2}(c) shows phase-resolved $\mu$-BLS scans across a narrow head-to-tail (red curve) and broad head-to-head (black curve) domain wall. The line scans were performed along the orange arrow in Fig. \ref{fig:2}(a). Since all optical conditions were maintained during the experiments, the spin-wave intensities for the two domain-wall configurations can be compared directly. While the signal intensity drops significantly behind the narrow head-to-tail domain wall, a similar suppression is not measured after the magnetization is switched to a broad head-to-head wall. For the narrow head-to-tail wall, the quick decay of the phase-sensitive $\mu$-BLS signal indicates strong reflection of propagating spin waves by the pinned domain wall.  

\subsection*{Micromagnetic simulations}

\begin{figure}[ht]
	\centering
	\includegraphics[width=\linewidth]{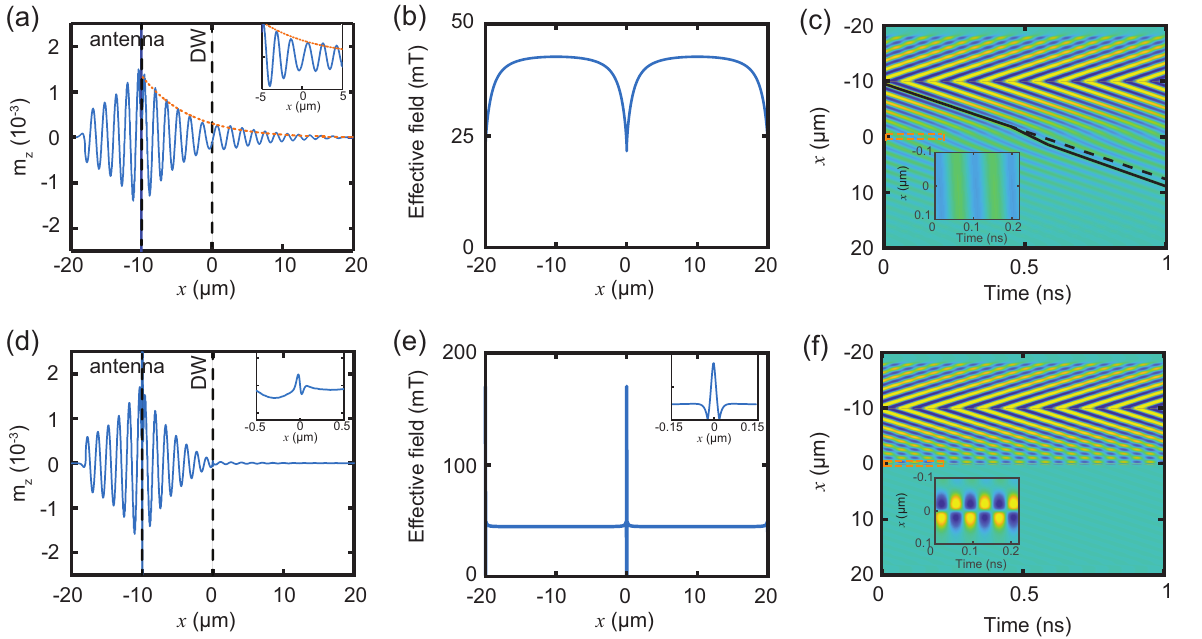}
	\caption{Micromagnetic simulations of spin-wave transport through a broad 90$^\circ$ head-to-head domain wall (a,c) and a narrow 90$^\circ$ head-to-tail domain wall (d,f) at a frequency of 10.85 GHz. The panels show data for the steady excitation state which is reached 9 ns after the sinusoidal magnetic field is turned on. The effective magnetic fields of the two domain walls are plotted in (b) and (e). The dashed line in (a) illustrates the amplitude of propagating spin waves if no domain wall is present. The inset in (f) illustrates strong out-of-phase oscillations of two anti-nodes inside the head-to-tail domain wall.}
	\label{fig:3}
\end{figure}

We performed micromagnetic simulations in MuMax3\cite{VAN-14} to further analyze spin-wave transmission through 90$^\circ$ N\'{e}el walls. In the simulations, we considered  two 20-$\mu$m-wide stripe domains. We used two-dimensional periodic boundary conditions and added a 1~$\mu$m wide region with higher damping parameter at the edges of the simulation area to prevent spin-wave interference. The structure was discretized using finite-difference 2.4~nm $\times$ 9.6~nm $\times$ 12.5~nm cells. To mimic anisotropy modulations in the experimental system, we abruptly rotated the uniaxial anisotropy axis by 90$^\circ$ at the domain boundary. Input parameters were derived from experiments. We extracted $M_s$ = 1.15$\times10^{6}$~A/m from vibrating sample magnetometry and $K_u$ = 2.5$\times10^{4}$~J/m$^3$ from BLS for our 50-nm-thick CoFeB film on BaTiO$_3$. Additionally, we used an exchange constant of $A_{ex}$ = 2.1$\times10^{-11}$~J/m and a damping parameter of $\alpha$ = 0.005. For these experimental parameters, we simulated a domain wall width of 50 nm (head-to-tail wall) and 1.6 $\mu$m (head-to-head and tail-to-tail wall). Spin waves were excited locally by an out-of-plane sinusoidal magnetic field at the center of one of the domains, i.e., 10 $\mu$m from the pinned domain wall. More details on micromagnetic simulations are given in the Methods section. 

Figure \ref{fig:3} summarizes simulation results for spin-wave transmission through a 90$^\circ$ head-to-head (a-c) and 90$^\circ$ head-to-tail (d-e) domain wall at a frequency of 10.85 GHz. In the head-to-head configuration, the effective magnetic field reduces gradually inside the domain wall (Fig. \ref{fig:3}(b)) and the magnetization rotates slowly more perpendicular to the wave vector of the propagating spin waves (Fig. \ref{fig:1}). Both effects change the spin-wave dispersion relation. While a decrease of effective field shifts the dispersion curve down, magnetization rotation towards the Damon-Eshbach configuration increases its slope and, thereby, the spin-wave group velocity. At constant excitation frequency, these local changes of the dispersion relation do also modify the spin-wave wavelength. As illustrated by the varying slope in the contour plot of Fig. \ref{fig:3}(c), the changes in group velocity and wavelength produce a finite phase shift when spin waves pass the head-to-head domain wall. The enhanced group velocity also reduces the decay of the spin-wave amplitude. This effect is illustrated by the dashed line in Fig. \ref{fig:3}(a), showing the envelope function of propagating spin waves for an uniform domain without magnetic domain wall. Thus, the broad 90$^\circ$ head-to-head domain wall is fully transparent to spin waves and, upon transmission, it alters the spin-wave phase and limits their attenuation.   

In contrast, spin waves are strongly reflected by the narrow 90$^\circ$ head-to-tail domain wall (Figs. \ref{fig:3}(d) and \ref{fig:3}(f)). The effective magnetic field inside this domain wall peaks sharply at its center (Fig. \ref{fig:3}(e)). Two field minima at $x=\pm$20 nm surround this peak. The non-uniform field profile produces a resonance mode that is characterized by two oscillatory out-of-phase antinodes on opposite sides of the domain-wall center (see inset in Fig. \ref{fig:3}(f) and Supplementary movie). Propagating spin waves are absorbed and reflected by this domain-wall resonance mode, as illustrated by the undulations of spin-wave maxima and minima in the contour graph of Fig. \ref{fig:3}(f). The undulations are produced by interference of counter-propagating spin waves. Because of interference, the spin-wave amplitude in front of the head-to-tail domain wall (Fig. \ref{fig:3}(d)) is reduced in comparison to the head-to-head configuration (Fig. \ref{fig:3}(a)). Similar resonance modes have been simulated previously for 180$^\circ$ domain walls in magnetic nanowires and it was shown that oscillating antinodes trigger spin-wave emission when driven by an external magnetic field\cite{HER-09} or cause resonant reflection of incoming spin waves\cite{WAN-APL-13}.   

\begin{figure}[ht]
	\centering
	\includegraphics[width=\linewidth]{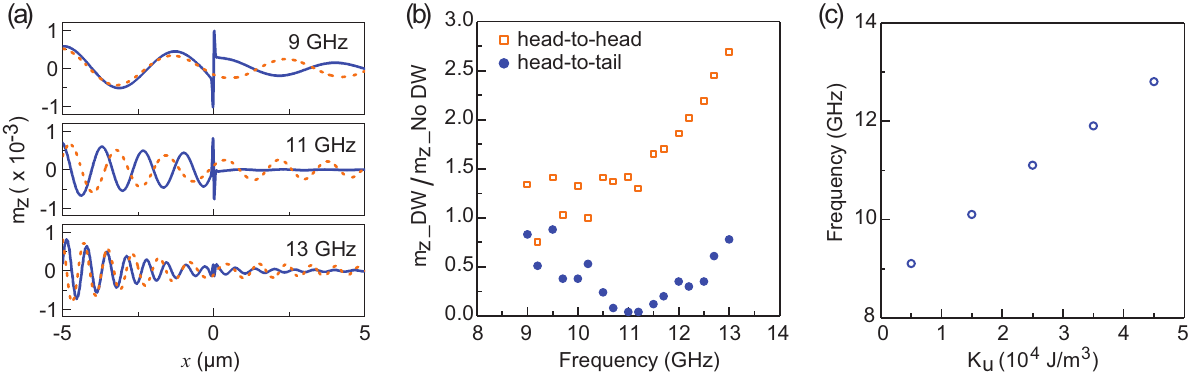}
	\caption{(a) Micromagnetic simulations of spin-wave transport through a broad 90$^\circ$ head-to-head domain wall (dashed orange line) and a narrow 90$^\circ$ head-to-tail domain wall (solid blue line) at three different frequencies. (b) Frequency dependence of the spin-wave amplitude after passing through a head-to-head and head-to-tail domain wall. The amplitude is normalized to simulation data for a film without domain wall. (c) Frequency of minimal spin-wave transmission through a head-to-tail domain wall as a function of magnetic anisotropy strength.}
	\label{fig:4}
\end{figure}

Figure \ref{fig:4} shows the frequency dependence of the spin-wave filtering effect. In (b), we plot the amplitude of spin waves after passing a magnetic domain wall, relative to their amplitude in the same film without domain wall. Obviously, the 90$^\circ$ head-to-tail wall reduces the spin-wave amplitude in the frequency range from 9 GHz to 13 GHz, with almost zero transmission around 11 GHz. The group velocity enhancement in the broad head-to-head domain wall, on the other hand, reduces the attenuation of spin waves. The frequency dependence of $m_\mathrm{z, DW}/m_\mathrm{z, No DW}$ for this domain wall is qualitatively explained by the variation of $m_\mathrm{z}$ with group velocity ($v_\mathrm{g}$) and spin-wave scattering time ($\tau$), $m_\mathrm{z}(x)\propto\mathrm{exp}(-x/v_\mathrm{g}\tau)$. Since the spin-wave scattering time is approximated by $\tau=1/2\pi\alpha{f}$\cite{MAD-11}, we can write $m_\mathrm{z, DW}/m_\mathrm{z, No DW}\propto\mathrm{exp}(2\pi{x}\alpha{f}(1/v_\mathrm{g, No DW}-1/v_\mathrm{g, DW})$. Since the spin-wave group velocity is enhanced inside the head-to-head domain wall ($v_\mathrm{g, DW}>v_\mathrm{g, No DW}$), this expression predicts an exponential increase of $m_\mathrm{z, DW}/m_\mathrm{z, No DW}$ with frequency.  

\section*{Magnetic spin-wave valve}

\begin{figure}[ht]
	\centering
	\includegraphics[width=\linewidth]{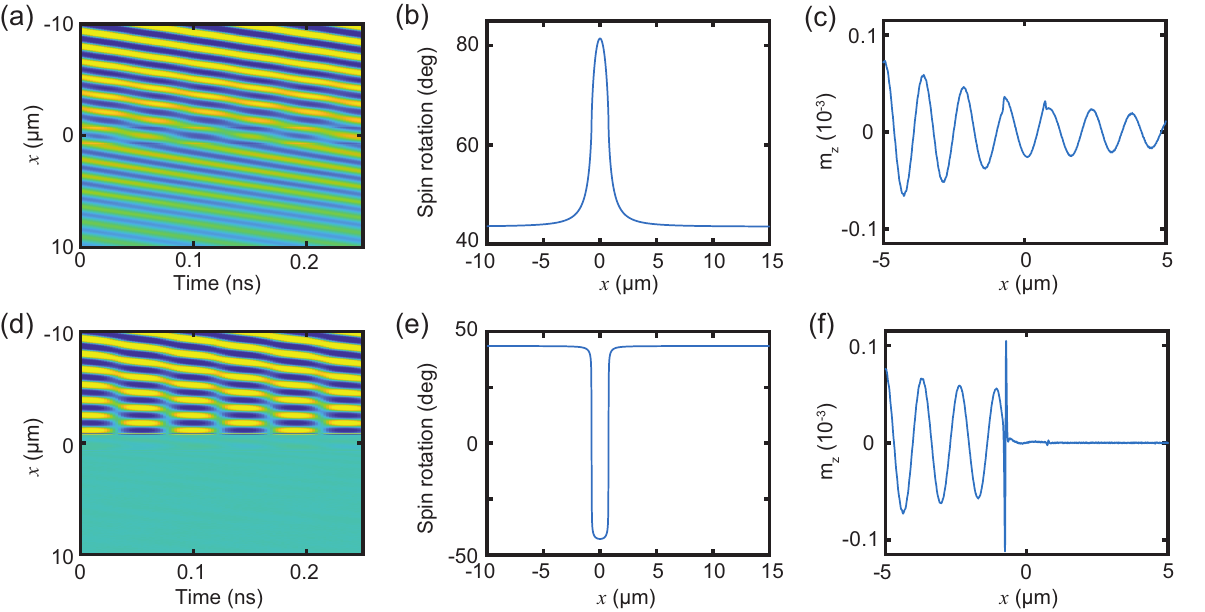}
	\caption{Micromagnetic simulations of spin-wave transport through a magnetic spin-wave valve with two pinned domain walls. The distance between the walls is 1.5~$\mu$m and the frequency is 11 GHz. Spin waves are excited at $x$ = -10 $\mu$m and the panels show data for steady-state excitation. Magnetization reversal in the central domain transforms a head-to-head/tail-to-tail wall combination (transport characteristics in (a) and (c)) into two head-to-tail domain walls (transport results in (d) and (f)), or vice versa. Spin rotations between domains for these two states are shown in (b) and (e).}
	\label{fig:5}
\end{figure}

Following our results on spin-wave transmission through single 90$^\circ$ domain walls, we propose a new structure for active spin-wave manipulation. Our device concept consists of three stripe domains with uniaxial magnetic anisotropy and two pinned domain walls (Fig. \ref{fig:1}(a)). In this configuration, magnetization reversal in the central domain switches the domain-wall state between a head-to-head/tail-to-tail combination and two head-to-tail walls. In practice, toggling between these two remanent magnetization states can be achieved by applying a magnetic-field pulse along the anisotropy axis of the central domain. Switching broad domain walls into narrow domain walls dramatically changes the transmission of spin waves near the domain-wall resonance frequency. An example at $f=11$ GHz is shown in Fig. \ref{fig:5}. The pinned domain walls are separated by 1.5~$\mu$m in this simulation. For the narrow head-to-tail walls, this distance is sufficient to reach a spin rotation of 90$^\circ$ (Fig. \ref{fig:5}(b)). In contrast, the spin rotation is only $38^\circ$ for the structure with a head-to-head/tail-to-tail wall combination. Irrespective of this finite-size scaling effect\cite{FRA-14}, the configuration with broad domain walls is fully transparent for propagating spin waves (Figs. \ref{fig:5}(a) and \ref{fig:5}(c)). The excitation of a resonance mode in the narrow domain walls, on the other hand, reduces spin-wave transmission to nearly 0\% (Figs. \ref{fig:5}(d) and \ref{fig:5}(f)). Since the spin-wave signal can be easily turned on or off by magnetic switching of one single domain, we refer to our structure as a magnetic spin-wave valve. 

\section*{Discussion}

Just like reconfigurable magnonic crystals\cite{KRA-14,CHU-17,TOP-10}, properties of the magnetic spin-wave valve can be changed on demand, enabling its application as magnon conduit, filter, or logic gate. Moreover, since the on and off states are non-volatile, the same element can be used to store data. Downscaling of a three-domain element is limited by the width of the head-to-tail domain wall. For typical materials and uniaxial anisotropy strengths, this type of domain wall is several tens of nanometers wide. At such length scales, the on state does no longer contain pinned magnetic domain walls (the width of head-to-head/tail-to-tail domain walls is too large to 'fit' into nanoscale domains). Instead, the magnetization is nearly uniform with negligible spin rotation between domains\cite{FRA-14}. Spin waves propagate through this magnetization configuration without significant pertubation, enabling deterministic switching between the on (no domain walls) and off (two narrow head-to-tail domain walls) states. To realize switching in patterned devices, one could use local Oersted fields from a metallic nanowire that is placed on top of the ferromagnetic film. 

The proposed magnetic spin-wave valve requires a regular modulation of magnetic anisotropy. Here, we used strain coupling in a CoFeB/BaTiO$_3$ bilayer to demonstrate the operation principle. There are, however, other means by which lateral changes in magnetic anisotropy can be realized. Examples include area-selective ion irradiation\cite{TRU-14,TRU-16} and thermally-assisted scanning probe lithography\cite{ALB-16}. Fabrication of a magnetic spin-wave valve is thus not limited to the use of specific substrates. 

For the experimental parameters of our system, the transmission of spin waves reduces drastically by resonant reflection around 11 GHz. In analogy to magnonic crystals, we estimate a spin-wave bandgap from the transmission curve (Fig. \ref{fig:4}(b)). Using the full width at half maximum, we find a bandgap of $\Delta{f}\approx$ 2 GHz. The domain-wall resonance mode and, thereby, the magnetic spin-wave valve operation frequency can be tuned by variation of the anisotropy strength (Fig. \ref{fig:4}(c)). A lowering of the uniaxial magnetic anisotropy reduces the frequency of minimal spin-wave transmission. In our strain-coupled system, this could be achieved by a change of the deposition parameters, the insertion of a seed layer at the ferromagnetic/ferroelectric interface, or the use of a ferromagnetic material with a smaller magnetostriction constant.           

In summary, we report on active spin-wave manipulation using a domain-wall-based magnetic spin-wave valve. Control over the transmission of propagating spin waves is realized by magnetic-field induced transformation of two pinned domain walls. If the spin-wave valve is set to comprise broad head-to-head and tail-to-tail walls, the transmission coefficient approaches 100\%. Switching to a state with two narrow head-to-tail walls reduces the transmission to nearly 0\% at the frequency of a domain-wall resonance mode. Both magnetization configurations are non-volatile and toggling between the two states is fully reversible and easily attained by magnetic switching in the central domain of the spin-wave valve. 

\section*{Methods}

\subsection*{Sample preparation}

We grew the 50-nm-thick CoFeB film with a composition of 40\% Co, 40\%Fe, and 20\% B on a single-crystal BaTiO$_3$ substrate using dc magnetron sputtering at 175 $^\circ$C. At this temperature, BaTiO$_3$ is paraelectric and its lattice exhibits cubic symmetry. During post-deposition cooling through the paraelectric-to-ferroelectric phase transition at 120 $^\circ$C, the structure of BaTiO$_3$ becomes tetragonal and a regular pattern of ferroelectric stripe domains forms. The polarization in the domains is oriented in-plane and, together with the elongated axis of the tetragonal unit cell, it rotates by 90$^\circ$ at domain boundaries. The phase transition in the BaTiO$_3$ substrate strains the CoFeB film and, via inverse magnetostriction, this imposes regular rotations of the uniaxial magnetic anisotropy axis. Since domain walls are strongly pinned by the magnetic anisotropy boundaries, it is possible to control their spin structure by the application of a magnetic field\cite{FRA-12}. After cooling to room temperature, we covered the CoFeB film by a 3 nm Ta/28 nm TaO$_x$ bilayer. The TaO$_x$ film was grown by reactive sputtering. Microwave antennas with a width of 500 nm were patterned onto the TaO$_x$ film using electron-beam lithography and lift-off. The broadband antennas consisted of 3 nm Ta and 50 nm Au. 

\subsection*{Magnetic characterization and Brillouin light scattering experiments}

The magnetic domain structure of the sample was imaged using a wide-field magneto-optical Kerr microscope with 20$\times$ and 100$\times$ objectives. We used vibrating sample magnetometry to measure the saturation magnetization of the CoFeB film. Phase-resolved $\mu$-BLS measurements were performed using a 100$\times$ objective near the center of one of the excitation antennas and along the direction of propagating spin waves. We selected an antenna with a pinned domain wall in its vicinity. To maintain positional accuracy, we supplied the same reference image as feedback to an image recognition-based drift stabilization software in experiments on both head-to-head and head-to-tail domain walls. More details on the $\mu$-BLS setup can be found in Ref.~\citenum{MAD-12}. 

\subsection*{Micromagnetic simulations}

We performed micromagnetic simulations using open-source GPU-accelerated MuMax3 software. Before spin-wave excitation, the simulation geometry was initialized by aligning the magnetization along the uniaxial anisotropy axes and letting the system reach its ground state in zero magnetic field. After selection of the domain-wall type, spin waves were excited locally by a 100 mT out-of-plane sinusoidal magnetic field. The ac field was applied to a one-cell-wide line at the center of one of the domains. To visualize propagating spin waves, we recorded the z-component of magnetization after reaching steady-state excitation. It took about 9 ns to reach this state. 


\section*{Acknowledgements}

This work was supported by the European Research Council (Grant No. ERC-2012-StG 307502-E-CONTROL). S.J.H. acknowledges financial support from the V\"ais\"al\"a Foundation. Lithography was performed at the Micronova Nanofabrication Centre, supported by Aalto University. We also acknowledge the computational resources provided by the Aalto Science-IT project.

\section*{Author contributions statement}

S.J.H. and S.v.D. designed and initiated the research. S.J.H. fabricated the samples. S.J.H., M.M., and G.G. conducted the $\mu$-BLS measurements. S.J.H. and H.J.Q. performed the micromagnetic simulations. S.v.D. supervised the project. S.J.H. and S.v.D. wrote the manuscript, with input from all other authors.  

\end{document}